\documentclass[pdflatex,sn-mathphys-num]{sn-jnl}

\usepackage{graphicx}%
\usepackage{multirow} 
\usepackage{multirow}%
\usepackage{amsmath,amssymb,amsfonts}%
\usepackage{amsthm}%
\usepackage{mathrsfs}%
\usepackage[title]{appendix}%
\usepackage{xcolor}%
\usepackage{textcomp}%
\usepackage{manyfoot}%
\usepackage{booktabs}%
\usepackage{algorithm}%
\usepackage{algorithmicx}%
\usepackage{algpseudocode}%
\usepackage{listings}%
\usepackage{booktabs}%
\usepackage{threeparttable}
\usepackage{float} 

\theoremstyle{thmstyleone}%
\theoremstyle{thmstyletwo}%

\theoremstyle{thmstylethree}%

\begin{document}

\title[Article Title]{Global impacts of organic aerosol acidity on sulfate and cloud formation}



\author[1]{\fnm{Gargi} \sur{Sengupta}}

\author[1]{\fnm{Kunal} \sur{Ghosh}}

\author[1]{\fnm{Prithvi R.} \sur{Jallu}}

\author*[1,2,3]{\fnm{N{\o}nne L.} \sur{Prisle}}\email{noenne.prisle@desy.de, nonne.prisle@oulu.fi}

\affil[1]{\orgdiv{Center for Atmospheric Research}, \orgname{University of Oulu}, \orgaddress{\street{P.O. Box 4500}, \city{Oulu}, \postcode{90014}, \country{Finland}}}

\affil[2]{\orgdiv{Center for Molecular Water Science}, \orgname{Deutsches Elektronen-Synchrotron DESY}, \orgaddress{\street{Notkestrasse 85}, \postcode{D-22607} \city{Hamburg}, \country{Germany}}}

\affil[3]{\orgdiv{Institute of Inorganic and Applied Chemistry}, \orgname{University of Hamburg}, \orgaddress{\street{Martin-Luther-King-Platz 6}, \city{Hamburg}, \postcode{D-20146}, \country{Germany}}}

\maketitle



\abstract{Organic aerosols (OA) comprise a major fraction of atmospheric particulate matter and frequently contain acidic species, yet their contribution to overall aerosol acidity has not been explicitly considered in global climate models.
We implement concentration-dependent OA acid dissociation, including recently demonstrated surface-specific effects, into the ECHAM-HAMMOZ global climate model and assess the impacts on aqueous aerosol sulfate chemistry and aerosol--cloud--climate interactions. 
We show that enhanced aerosol acidity from OA acid dissociation drives increased sulfate formation from aqueous-phase oxidation of $\mathrm{SO_2}$. 
The microphysics of additional secondary sulfate aerosol changes global cloud droplet number concentrations (CDNC), with enhancements up to $13.9\%$. 
Increased 
cloud formation leads to a significant global mean cooling effect with a shortwave cloud radiative forcing (SWCRF) up to $-0.97~\mathrm{W\,m^{-2}}$.
We also find that surface-specific acid dissociation effects can further modify both aerosol chemistry and resulting aerosol--cloud--climate responses, in some cases with even stronger impact than bulk acidity conditions.
Our results demonstrate significant effects of considering OA acidity, as well as surface-specific phenomena, in global climate models.





}





\section{Introduction}\label{sec1}

Atmospheric aerosols play a significant role in the global climate system, directly influencing Earth's radiation budget and further acting as cloud condensation nuclei (CCN), thereby affecting cloud formation, properties, and climate feedbacks \cite{forster:ipcc:2021, stocker:ipcc:2013}. 
Organic compounds are prevalent in atmospheric aerosols and can comprise $20-50\%$ of aerosol mass in mid-latitude regions and up to $90\%$ in tropical forests \cite{andreae:science:1997, putaud:atmenv:2004, kanakidou:acp:2005}. 
Despite their abundance, 
organic aerosols (OA) 
still contribute substantial uncertainties in climate models and their predictions \cite{stocker:ipcc:2013}.


Organic aerosols frequently contain Br{\o}nsted acidic or basic compounds \cite{Jacob:jgratmos:1986, Millet:acp:2015, Chen:esnt:2021, Angelis:chemphysmet:2012, Grisillon:talanta:2026}, particularly carboxylic acids, which can account for up to $12-32\%$ of the water-soluble organic fraction in marine and continental aerosols \cite{chebbi:atmenv:1996} and approximately $6-15\%$ of total OA mass in urban environments \cite{Wu:waterairsoil:2020}. 
These components 
contribute to overall aerosol acidity by influencing the concentration of solvated hydrogen ions \cite{pye:acp:2020, ault:accounts:2020}, which in turn regulates the protonation state of all aerosol components. 
This has important implications for aerosol chemistry, including water activity and aqueous-phase formation of secondary aerosol mass \cite{yli:acp:2013, prisle:chemphysmet:2008}.  
However, global climate models currently lack detailed representation of OA acidity \cite{kanakidou:acp:2005, prisle:acp:2012, pye:acp:2020}, including explicit consideration of concentration-dependent acid dissociation. 

Many atmospheric organic acids and bases, for example mono- and dicarboxylic acids, organosulfates, and alkyl amines, also exhibit surface activity in aqueous solutions
\cite{mochida:jgratmospheres:2002, mochida:jgratmospheres:2003, prisle:acp:2010, ottosson:pccp:2011, ohrwall:jphyschem:2015, hansen:acp:2015, hyttinen:acp:2020, prisle:accounts:2024}. 
Surface-active species (often called "surfactants") have been reported in atmospheric OA from many different regions and environments \cite{petters:jgratmospheres:2016, noziere:jove:2017, kroflic:esnt:2018, gerard:esnt:2019}. 
Surfactants adsorb at the aqueous surface, leading to enhanced concentrations compared to the interior (bulk). 
In nano- and microscopic atmospheric aerosols and droplets with large surface-area-to-volume ratios, surface adsorption can result in significant bulk--surface partitioning and strong depletion of the bulk phase concentration \cite{prisle:acp:2010, prisle:acp:2021}. 
When a large fraction of the surface-active material resides in the surface, while the bulk is highly dilute, the surface chemical and physical state may play a critical role in determining the overall aerosol properties \cite{prisle:acp:2012, prisle:accounts:2024}.


Recent molecular-level experiments using X-ray photoelectron spectroscopy in combination with high-brilliance synchrotron radiation have shown that the protonation state of atmospheric organic acids and bases at the aqueous surface can differ strongly from the bulk solution \cite{prisle:acp:2012, ohrwall:jphyschem:2015, werner:pccp:2018, prisle:accounts:2024}. 
The apparent $\mathrm{pK_a}$ at the surface was observed to be shifted by $1-2$~$\mathrm{pH}$ units, indicating significant surface-specific suppression of OA acid dissociation. 
These observations are supported by $\mathrm{pH}$-dependent surface tension measurements of monocarboxylic acids \cite{wellen:pccp:2017} 
and 
molecular dynamics simulations showing weaker acid strength 
at the aqueous surface \cite{puente:jacs:2022}.

Aerosol acidity in global models is typically represented using thermodynamic modules
\cite{pye:acp:2020}. 
Chemical transport models (CTMs) 
calculate aerosol $\mathrm{pH}$ based on interactions between aerosol components and ambient conditions, including relative humidity, temperature, and gas--particle partitioning. 
A summary of common CTMs, $\mathrm{pH}$ calculation methods, the species included, and their sources, are presented by \citet[][Table 7] {pye:acp:2020}.
To date, global climate simulations have assumed that OA components do not dissociate in the aqueous phase and therefore do not contribute to overall aerosol acidity \cite{kanakidou:acp:2005, hennigan:acp:2015, pye:acp:2020}. 
Surface-specific properties of aerosols and droplets are also typically not considered in global climate models. A few works to date have evaluated surfactant effects on aerosol--cloud--climate interactions \cite{prisle:grl:2012}. 
However, neither of these considered the acidity of surface-active OA components.

Recently, we implemented a representation of OA acidity in the ECHAM6.3–HAM2.3 aerosol--chemistry--climate box model (HAMBOX) with explicit consideration of concentration-dependent acid dissociation of OA and a possible surface-specific shift in their protonation state \cite{sengupta:acp:2024, sengupta:asnt:2024}. 
We now introduce OA acidity to the full ECHAM-HAMMOZ global climate model. This allows us to explore, for the first time, effects of OA acid dissociation on aqueous aerosol chemistry, cloud formation, and radiative forcing, in the fully coupled climate system and at a global scale. 

\section{Results}\label{sec2}


OA acid dissociation is implemented into the ECHAM-HAMMOZ global climate model (ECHAM6.3.0–HAM2.3–MOZ1.0) following the methodology of \citet{sengupta:acp:2024, sengupta:asnt:2024}, including both well-known bulk-phase and recently demonstrated surface-specific protonation effects.
Acid dissociation increases hydrogen ion concentrations, which impacts secondary sulfate formation in aqueous aerosols through acid-catalyzed oxidation of $\mathrm{SO_2}$ by $\mathrm{H_2O_2}$ and $\mathrm{O_3}$, and modifies aerosol water activity. 
We performed global simulations with four scenarios: (a) a base case without OA acid dissociation ($\rm K_a = 0$), (b) concentration-dependent partial dissociation according to the bulk-phase acidity ($\mathrm{pK^{B}_a}$), and two surface-specific conditions representing (c) moderate ($\mathrm{pK^{S1}_a} = \mathrm{pK^{B}_a} + 1$) and (d) strong ($\mathrm{pK^{S2}_a} = \mathrm{pK^{B}_a} + 2$) suppression of OA acid dissociation.
The entire OA fraction is assumed to consist of surface-active organic acids, represented by decanoic acid, a common proxy for atmospheric OA \cite{khwaja:atmenv:1995, mochida:jgratmospheres:2002, prisle:acp:2010, prisle:acp:2012, vepsalainen:acp:2023}.
Global effects of OA acid dissociation are evaluated in terms of sulfate aerosol burden, cloud droplet number concentration, cloud liquid water content, and resulting shortwave cloud radiative effect and forcing.

\subsection{Sulfate burden}\label{sec-su}

\begin{figure}[H]
  \centering
     \includegraphics[width=\textwidth]{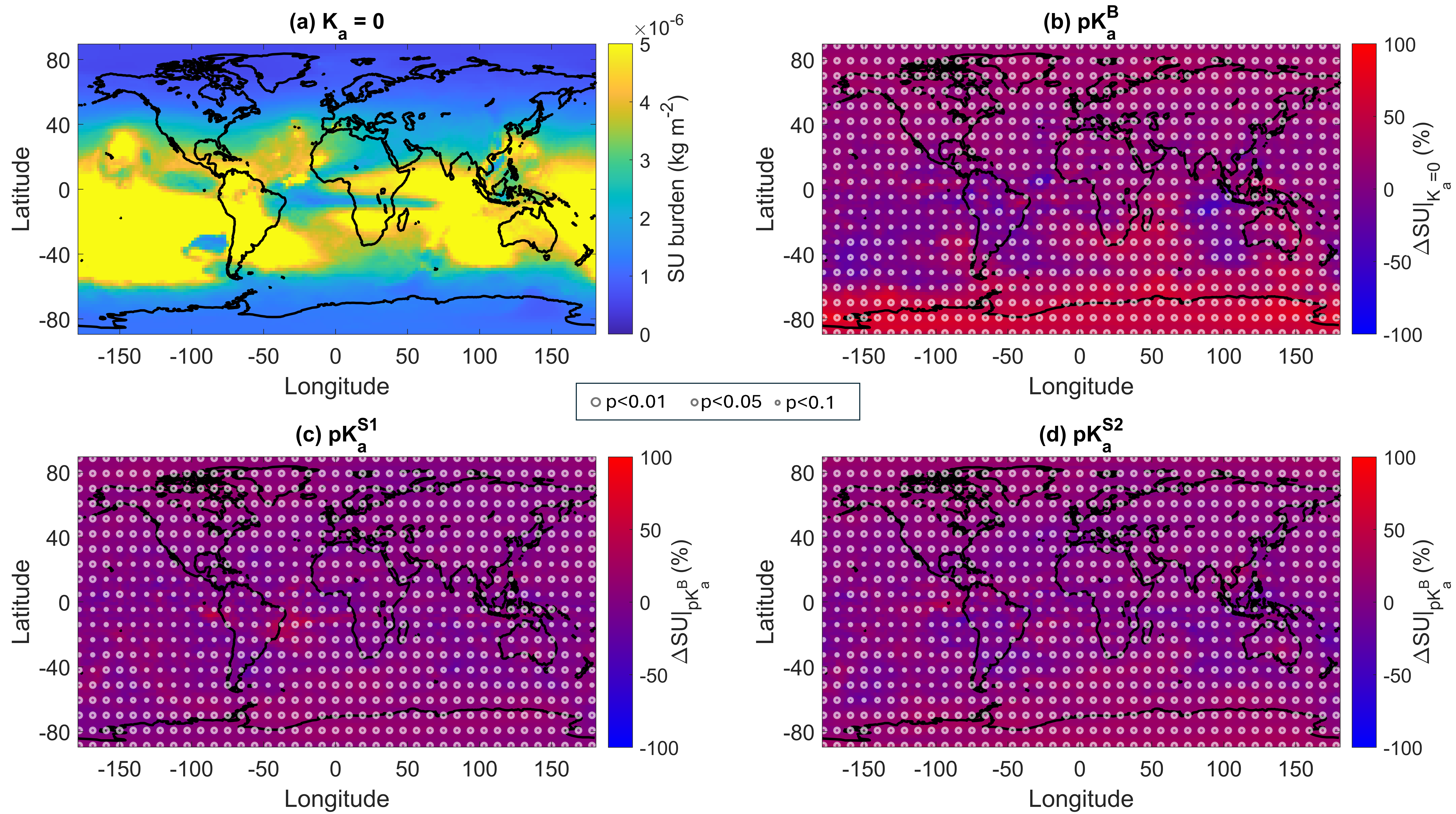}
      \caption{Sulfate aerosol (SU) burden shown as (a) the total column burden for the no OA acid dissociation condition ($\rm K_a=0$) as 5-year ($1999-2003$) medians in absolute units (\(\rm kg \, m^{-2}\)), and the column burden differences (b) with respect to $\rm K_a=0$ ($\Delta \mathrm{SU} \big|_{\mathrm{K_{a}=0}}$, \%) for bulk OA acid dissociation ($\mathrm{pK^{B}_a}$), and with respect to $\mathrm{pK^{B}_a}$ ($\Delta \mathrm{SU} \big|_{\mathrm{pK_a^B}}$, \%) for the surface-specific OA acid dissociation (c) $\mathrm{pK^{S1}_a}$ and (d) $\mathrm{pK^{S2}_a}$. Statistical significances of differences given as $p < 0.01$ (strong), $p < 0.05$ (medium), or $p < 0.1$ (weak) are represented by large, medium, and small circles, respectively.}
  \label{fig:sulf}
\end{figure}


Figure~\ref{fig:sulf} presents the global sulfate (SU) burden as 5-year medians from \(1999\)--\(2003\). 
The absolute SU burden ($\rm kg \, m^{-2}$) predicted for the no dissociation ($\rm K_a = 0$) condition is shown in panel (a), the change $\Delta \mathrm{SU} \big|_{\mathrm{K_{a}=0}}$ (\%) for bulk dissociation ($\mathrm{pK_a^B}$) relative to \(\rm K_a=0\) in panel (b), and \(\Delta \mathrm{SU} \big|_{\mathrm{pK_a^B}}\) (\%) for surface-specific conditions ($\rm pK_a^{S1}$, $\rm pK_a^{S2}$) relative to $\mathrm{pK_a^B}$ in panels (c) and (d). The statistical significance is evaluated as strong ($p < 0.01$, large circles), moderate ($p < 0.05$, medium circles), or weak ($p < 0.1$, small circles), using the Wilcoxon signed-rank test (Section~\ref{Sec:sig}). Average SU column burdens for each OA acid dissociation condition, both globally and regionally over land and ocean areas, are summarized in Table~\ref{tbl:SU}.

\begin{table}[h]
\centering
\caption{5-year absolute median SU burden ($\times 10^{-6} \, \rm kg \, m^{-2}$) and relative changes (\%) in SU burden with respect to $\mathrm{K_{a}}=0$ ($\Delta \mathrm{SU} \big|_{\mathrm{K_{a}=0}}$) and $\mathrm{pK_a^B}$ ($\Delta \mathrm{SU} \big|_{\mathrm{pK_a^B}}$) for bulk ($\mathrm{pK_a^B}$) and surface-specific ($\mathrm{pK_a^{S1}}$, $\mathrm{pK_a^{S2}}$) OA acid dissociation over the entire globe (world), land, and ocean regions.} 
\label{tbl:SU}
\begin{tabular}{llccc}
\hline
Region & $\mathrm{pK_a}$ & SU Burden ($\times 10^{-6} \, \rm kg \, m^{-2}$) & $\Delta \mathrm{SU} \big|_{\mathrm{K_{a}=0}}$ (\%) & $\Delta \mathrm{SU} \big|_{\mathrm{pK_a^B}}$ (\%) \\ 
\hline
    World 
    & $\mathrm{K_{a}}=0$  & 3.32 & --    & --    \\
    & $\mathrm{pK_a^B}$ & 3.58 & 7.76  & --    \\
    & $\mathrm{pK_a^{S1}}$ & 3.64 & 9.41  & 1.53 \\
    & $\mathrm{pK_a^{S2}}$ & 3.68 & 10.65 & 3.07 \\
\hline
    Land  
    & $\mathrm{K_{a}}=0$  & 2.29 & --    & --    \\
    & $\mathrm{pK_a^B}$ & 2.66 & 16.40 & --    \\
    & $\mathrm{pK_a^{S1}}$ & 2.73 & 19.37 & 2.55 \\
    & $\mathrm{pK_a^{S2}}$ & 2.81 & 23.09 & 6.02 \\
\hline
    Ocean
    & $\mathrm{K_{a}}=0$  & 5.44 & --    & --    \\
    & $\mathrm{pK_a^B}$ & 5.46 & 0.37  & --    \\
    & $\mathrm{pK_a^{S1}}$ & 5.48 & 0.88  & 0.51 \\
    & $\mathrm{pK_a^{S2}}$ & 5.44 & 0.02  & -0.37 \\
\hline
\end{tabular}
\end{table}


Without OA acid dissociation, the global SU burden is $3.32 \times 10^{-6}\,\rm kg \, m^{-2}$, with highest values over industrial areas and downwind oceanic regions, especially in the Northern Hemisphere. Lower burdens are seen over remote regions with limited precursor emissions.
%
%
%
Bulk OA acid dissociation increases the SU burden by $7.76\%$ globally, with a stronger land response ($16.40\%$), particularly over Northern Hemisphere industrial areas ($p < 0.01$), and minimal change ($0.37\%$, $p < 0.05$) over oceans. 
Surprisingly, surface-specific OA acid dissociation leads to further increase in global SU burden, $\Delta \mathrm{SU} \big|_{\mathrm{pK_a^B}}=1.53\%$ ($\rm pK_a^{S1}$) and $3.07\%$ ($\rm pK_a^{S2}$), with greater increases ($2.55\%$ and $6.02\%$, respectively) over land and minimal or slightly negative ($-0.37\%$ for $\rm pK_a^{S2}$) changes over oceans. 
Strongly significant changes ($p < 0.01$) appear mainly over land in the Northern Hemisphere.
%
The sulfate distribution across aerosol size modes is provided in the Supplementary information (Table~\ref{tab:combined_modes_deltas}). 
For the most CCN relevant accumulation (dry particle radius $R_p=0.05{-}0.5\,\mu$m) and coarse ($R_p>0.5\,\mu$m) modes, sulfate mass fractions ($X_\mathrm{Sulfate}$) are comparable, whereas total aerosol number concentration ($N_\mathrm{Total}$) and mass fraction-weighted sulfate number concentrations ($N_\mathrm{Sulfate}$) are much higher in the accumulation mode.
For both size modes, $X_\mathrm{Sulfate}$, $N_\mathrm{Total}$, and $N_\mathrm{Sulfate}$ all increase for $\rm pK_a^{B}$ compared to $\mathrm{K_a = 0}$. However, for $\rm pK_a^{S1}$ and $\rm pK_a^{S2}$ compared to $\rm pK_a^{B}$, $X_\mathrm{Sulfate}$, $N_\mathrm{Total}$, and $N_\mathrm{Sulfate}$ decrease for the accumulation mode but further increase for the coarse mode, showing different impacts of each OA acid dissociation condition on the distribution of CCN relevant aerosol.

\subsection{Cloud droplet number concentrations and liquid water content}\label{sec3}

\begin{figure}[h]
\centering
\includegraphics[width=\textwidth]{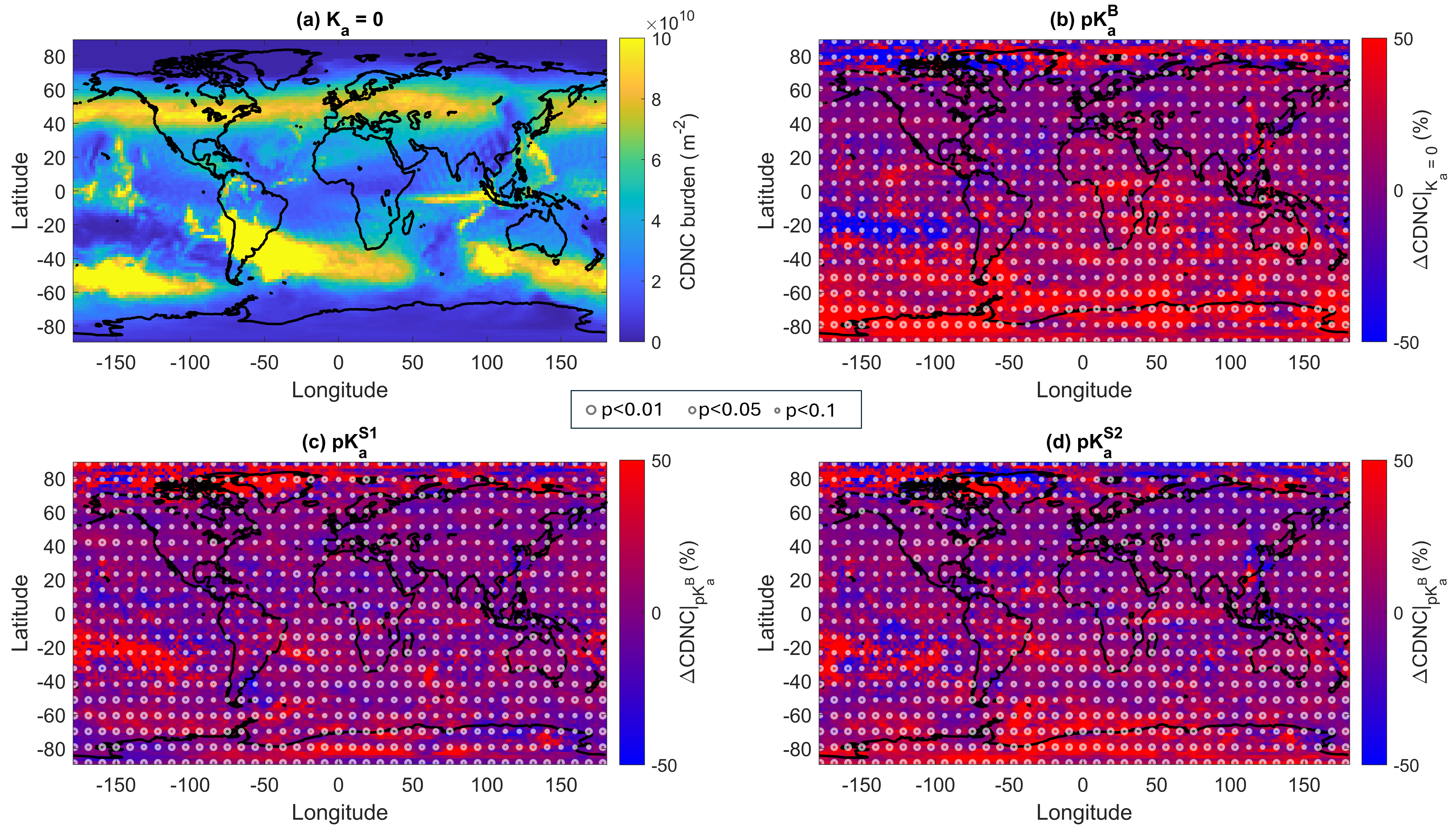}
\caption{CDNC burden shown as the total column CDNC for the no OA acid dissociation condition ($\rm K_a=0$, panel a) as 5-year ($1999-2003$) medians in absolute units ($\rm   m^{-2}$), and the column burden differences with respect to $\rm K_a=0$ ($\Delta \mathrm{CDNC} \big|_{\mathrm{K_{a}=0}}$, \%) for bulk OA acid dissociation ($\mathrm{pK^{B}_a}$, panel b), and with respect to $\mathrm{pK^{B}_a}$ ($\Delta \mathrm{CDNC} \big|_{\mathrm{pK_a^B}}$, \%) for the surface-specific OA acid dissociation  $\mathrm{pK^{S1}_a}$ (panel c) and $\mathrm{pK^{S2}_a}$ (panel d). Statistical significance (white circles) is indicated as large ($p < 0.01$), medium ($p < 0.05$), and small ($p < 0.1$).}\label{cdnc}
\end{figure}

Figure \ref{cdnc} shows the global distribution of total column cloud droplet number concentration (CDNC, \(\rm m^{-2}\)) as 5-year medians from \(1999\)--\(2003\). 
Panel (a) shows the absolute CDNC column burden for the condition of no OA acid dissociation (\(\rm K_a=0\)). 
The change \(\Delta \mathrm{CDNC} \big|_{\mathrm{K_{a}=0}}\) (\%) for the bulk acid dissociation condition $\mathrm{pK^{B}_a}$ relative to \(\rm K_a=0\) is shown in panel (b), and changes \(\Delta \mathrm{CDNC} \big|_{\mathrm{pK_a^B}}\) (\%) for surface-specific suppressed acid dissociation conditions $\mathrm{pK^{S1}_a}$ and $\mathrm{pK^{S2}_a}$, respectively, relative to $\mathrm{pK^{B}_a}$ in panels (c) and (d). 
Corresponding changes ($\Delta \mathrm{CDNC} \big|_{\mathrm{K_{a}=0}}$, \%) relative to $\mathrm{K_a = 0}$ are given in the Supplementary information (Fig.~\ref{cdncKa0}). 
Statistical significances in panels (b–d) 
are indicated as strong ($p < 0.01$), moderate ($p < 0.05$), or weak ($p < 0.1$) by large, medium, and small circles.
Regional CDNC burdens for each condition are summarized in Table \ref{tbl:cdnc}.

\begin{table}[h]
\centering
\caption{5-year absolute median CDNC ($\times10^{10} \, \mathrm{m^{-2}}$) and relative changes (\%) in CDNC with respect to $\mathrm{K_{a}}=0$ ($\Delta \mathrm{CDNC} \big|_{\mathrm{K_{a}=0}}$) and $\mathrm{pK_a^B}$ ($\Delta \mathrm{CDNC} \big|_{\mathrm{pK_a^B}}$) for bulk ($\mathrm{pK_a^B}$) and surface-specific ($\mathrm{pK_a^{S1}}$, $\mathrm{pK_a^{S2}}$) OA acid dissociation over the entire globe (world), land, and ocean regions.} 
\label{tbl:cdnc}
\begin{tabular}{llccc}
\toprule
Region & $\mathrm{pK_a}$ & CDNC ($\times10^{10} \, \mathrm{m^{-2}}$) & $\Delta \mathrm{CDNC} \big|_{\mathrm{K_{a}=0}}$ ($\%$) & $\Delta \mathrm{CDNC} \big|_{\mathrm{pK_a^B}}$ ($\%$) \\ 
\midrule
\multirow{4}{*}{World} 
    & $\mathrm{K_{a}}=0$  & $3.86$ & --    & --    \\
    & $\mathrm{pK_a^B}$ & $4.25$ & $10.15$ & --    \\
    & $\mathrm{pK_a^{S1}}$ & $4.31$ & $11.54$ & $1.41$ \\
    & $\mathrm{pK_a^{S2}}$ & $4.40$ & $13.92$ & $3.53$ \\
\midrule
\multirow{4}{*}{Land}  
    & $\mathrm{K_{a}}=0$  & $3.01$ & --    & --    \\
    & $\mathrm{pK_a^B}$ & $3.39$ & $12.68$ & --    \\
    & $\mathrm{pK_a^{S1}}$ & $3.43$ & $14.01$ & $1.18$ \\
    & $\mathrm{pK_a^{S2}}$ & $3.54$ & $17.48$ & $4.42$ \\
\midrule
\multirow{4}{*}{Ocean} 
    & $\mathrm{K_{a}}=0$  & $5.58$ & --    & --    \\
    & $\mathrm{pK_a^B}$ & $6.00$ & $7.37$  & --    \\
    & $\mathrm{pK_a^{S1}}$ & $6.08$ & $8.83$  & $1.33$ \\
    & $\mathrm{pK_a^{S2}}$ & $6.14$ & $10.01$ & $2.33$ \\
\bottomrule
\end{tabular}
\end{table}

When OA acid dissociation is not considered, 
high CDNC burdens are primarily observed over marine stratocumulus regions, such as the eastern Pacific and Atlantic Oceans off the coasts of South America and Africa, and in tropical convergence zones. Lower CDNC burdens are seen over much of the continental areas, especially over deserts and polar regions, which are typically aerosol-limited environments for cloud formation \cite{Mauritsen:acp:2011}. The spatial distribution of CDNC for the $\rm K_a = 0$ condition agrees with previous simulations using the same model setup and time period \citep{prisle:grl:2012}. 
The average column CDNC burden (Table \ref{tbl:cdnc}) is $3.86 \times 10^{10}\, \mathrm{m^{-2}}$ globally and slightly higher over oceans ($5.58 \times 10^{10}\, \mathrm{m^{-2}}$) than over land ($3.01 \times 10^{10}\, \mathrm{m^{-2}}$). 

Including OA bulk acid dissociation leads to a $10.15\%$ increase in global CDNC burden, reaching $4.25 \times 10^{10}~\mathrm{m^{-2}}$ (Table~\ref{tbl:cdnc}), with a stronger increase over land ($12.68\%$) than over oceans ($7.37\%$). 
This enhancement is primarily driven by additional secondary sulfate aerosol mass resulting from accelerated aqueous-phase oxidation of $\mathrm{SO_2}$ due to elevated hydrogen ion concentrations.
The magnitude of the CDNC response is consistent with our previous box model simulations using HAMBOX \cite{sengupta:acp:2024, sengupta:asnt:2024}, which showed increases of 
$\Delta \mathrm{CDNC} \big|_{\mathrm{K_{a}=0}}=5-50\%$ for varying environments (aerosol number concentrations of $350-3500~\times~10^6$~m$^{-3}$).
Despite the overall increase, some regions, particularly the tropics, show decreases in CDNC, demonstrating the simultaneous role of interactions with local ambient conditions. Regions with strongly significant changes ($p < 0.01$) are predominantly seen over oceans.
For the surface-specific conditions, global $\Delta \mathrm{CDNC} \big|_{\mathrm{pK_a^B}}=1.41\%$ ($\mathrm{pK_a^{S1}}$) and $3.53\%$ ($\mathrm{pK_a^{S2}}$), and increases are more pronounced over land ($1.18\%$ and $4.42\%$) than over oceans ($1.33\%$ and $2.33\%$). These changes are statistically significant in many regions, especially over continental areas.

There is a strong spatial variability in both absolute CDNC burden and the changes between conditions. 
The regional patterns in $\Delta \mathrm{CDNC} \big|_{\mathrm{K_{a}=0}}$ and $\Delta \mathrm{CDNC} \big|_{\mathrm{pK_a^B}}$ closely resemble those of the corresponding sulfate burden changes (Fig.~\ref{fig:sulf}, Table~\ref{tbl:SU}). 
For example, mid-latitude oceanic regions show strongly significant ($p < 0.01$) positive $\Delta \mathrm{CDNC} \big|_{\mathrm{K_{a}=0}}$, while tropical areas exhibit both positive and negative changes. 
Contrasts between areas with positive and negative $\Delta \mathrm{CDNC} \big|_{\mathrm{pK_a^B}}$ are more pronounced for $\mathrm{pK_a^{S2}}$ than for $\mathrm{pK_a^{S1}}$, with more and larger increases (red areas) at high latitudes and decreases (blue areas) in the tropics.
This suggests that effects of surface-specific suppressed OA acid dissociation interact more intricately with ambient conditions than those of the bulk acidity condition.

Cloud liquid water content (LWC) in particular plays an important role in shaping the CDNC response, together with sulfate burden and distribution across CCN relevant size modes. 
Results for LWC ($\rm kg \, m^{-2}$) are presented in the Supplementary information (Fig.~\ref{lwc}, Table~\ref{tbl:lwc}). 
In the surface dissociation condition $\mathrm{pK_a^{S2}}$, CDNC increases by $17.48\%$ (Fig. S1 and Table~\ref{tbl:cdnc}), while SU over land increases by $23.09\%$ relative to $\mathrm{K_a}=0$ (Fig.~\ref{fig:sulf}, Table~\ref{tbl:SU}) and LWC by only $9.95\%$. 
This suggests that limited cloud water availability constrains cloud droplet activation of additional secondary sulfate aerosol. Over oceans, where the $\mathrm{pK_a^{S2}}$ condition yields only a marginal sulfate increase ($\Delta \mathrm{SU} \big|_{\mathrm{K_{a}=0}}=0.02\%$), CDNC still increases by $10.01\%$ as LWC increases by $4.59\%$, each with respect to the $\mathrm{K_a = 0}$ condition. 
CDNC in marine regions is therefore more strongly limited by cloud water than by sulfate aerosol mass.
Over land, CDNC responds less strongly to additional sulfate, unless LWC also increases. 
Thus, while even small increases in sulfate aerosol can enhance CDNC when sufficient LWC is present, regions with limited LWC may not exhibit the same sensitivity. 
The limiting role of LWC is consistent with our previous results from box model simulations \cite{sengupta:asnt:2024}, where a fixed LWC of $0.03~\mathrm{g~m^{-3}}$ led to saturation effects at high aerosol concentrations, reducing CDNC enhancement from increased aerosol acidity and sulfate formation.
In the fully coupled climate model, low LWC arises from ambient conditions rather than imposed boundary conditions, but with similar effect. 



\subsection{Shortwave cloud radiative effect and radiative forcing}

\begin{table}[h]
\centering
  \caption{Shortwave cloud radiative effect (SWCRE, ~$\rm{W  \, m^{-2}}$) as 5-year ($1999-2003$) means, with the corresponding change ($\Delta \mathrm{SWCRE} \big|_{\mathrm{K_{a}=0}}$, \%) relative to $\rm K_a=0$ in parentheses for each of the OA acid dissociation conditions $\rm K_a=0$ (no OA acid dissociation), $\mathrm{pK^{B}_a}$ (bulk OA acid dissociation), and $\mathrm{pK^{S1}_a}$ and $\mathrm{pK^{S2}_a}$ (surface-specific OA acid dissociation), evaluated globally (world), and for land and ocean areas. Furthermore, 5-year mean shortwave cloud radiative forcing (SWCRF,~$\rm{W  \, m^{-2}}$) for each OA acid dissociation condition.} 
  \label{tbl:SWCRE}
  \begin{tabular}{lcccc}\hline
SWCRE &   $\rm{K_{a}}=0$    & $\mathrm{pK^{B}_a}$      & $\mathrm{pK^{S1}_a}$ &$\mathrm{pK^{S2}_a}$  \\ 
\hline
World & $-38.59$ & $-39.57$ ($-0.98$) & $-39.71$ ($-1.12$) & $-39.77$ ($-1.18$) \\ 
Land  & $-33.63$ & $-34.63$ ($-1.0$) & $-34.82$ ($-1.19$) & $-34.86$ ($-1.23$) \\ 
Ocean & $-48.68$ & $-49.62$ ($-0.94$)  & $-49.63$ ($-0.95$)  & $-49.76$ ($-1.08$) \\ \hline
SWCRF & $-0.35$ & $-0.65$  & $-0.79$  & $-0.97$ \\ \hline
  \end{tabular}
\end{table}

The 5-year ($1999-2003$) mean shortwave cloud radiative effect (SWCRE, in $\mathrm{W \, m^{-2}}$) and the relative changes ($\Delta \mathrm{SWCRE} \big|_{\mathrm{K_{a}=0}}$, in \%) with respect to $\mathrm{K_a = 0}$ for different conditions of OA acid dissociation are given in Table \ref{tbl:SWCRE}. 
Including OA acid dissociation leads to an enhanced cooling effect from 
cloud formation due to additional sulfate aerosol. 
Globally, SWCRE becomes increasingly negative, indicating a stronger shortwave cloud radiative cooling, when OA acidity is considered, and even more so for surface-specific dissociation, with $\Delta \mathrm{SWCRE} \big|_{\mathrm{K_{a}=0}}=-0.98\%$ ($\mathrm{pK^{B}_a}$), $-1.12\%$ ($\mathrm{pK^{S1}_a}$), and $-1.18\%$ ($\mathrm{pK^{S2}_a}$).
The total cooling effect is stronger over ocean regions, where SWCRE~$= -48.68 \, \mathrm{W \, m^{-2}}$ ($\mathrm{K_{a}}=0$) further decreases by $-0.94\%$ ($\mathrm{pK^{B}_a}$), $-0.95\%$ ($\mathrm{pK^{S1}_a}$), and $-1.08\%$ ($\mathrm{pK^{S2}_a}$). 
Over land areas, the additional cooling from OA acidity is greater, with $\Delta \mathrm{SWCRE} \big|_{\mathrm{K_{a}=0}}=-1.0\%$ ($\mathrm{pK^{B}_a}$), $-1.19\%$ ($\mathrm{pK^{S1}_a}$), and $-1.23\%$ ($\mathrm{pK^{S2}_a}$).
These changes are consistent with the $\Delta \mathrm{CDNC}$ patterns (Figure \ref{cdnc}, Table \ref{tbl:cdnc}), which also show stronger responses to additional secondary sulfate from OA acid dissociation over land, compared to ocean regions.

The 5-year mean shortwave cloud radiative forcing (SWCRF),
calculated as the change in SWCRE relative to clean, pre-industrial conditions,  
is also given in Table~\ref{tbl:SWCRE}.
Overall, the consideration of OA acid dissociation yields a non-negligible contribution to anthropogenic aerosol forcing in the present-day atmosphere. The cooling effect is further amplified for surface-specific conditions, compared to bulk OA acidity, with SWCRF~$=-0.35~\rm{W \, m^{-2}}$ ($\rm{K_{a}}=0$), $-0.65~\rm{W \, m^{-2}}$ ($\mathrm{pK^{B}_a}$), $-0.79~\rm{W \, m^{-2}}$ ($\mathrm{pK^{S1}_a}$), $-0.97~\rm{W \, m^{-2}}$ ($\mathrm{pK^{S2}_a}$).

\section{Discussion}\label{sec12}

We here present the first global-scale investigation of organic aerosol acid dissociation effects on aerosol and cloud formation and climate impact.
Concentration-dependent OA acid dissociation increases aqueous aerosol-phase hydrogen ion concentrations, promoting acid-catalyzed oxidation of $\mathrm{SO_2}$ by $\mathrm{H_2O_2}$ and 
$\mathrm{O_3}$. This leads to enhanced sulfate burden and cloud droplet number concentrations from secondary sulfate aerosol formation, particularly over marine and highly industrialized land regions. 
Our previous box model simulations~\cite{sengupta:acp:2024} showed that sulfate from $\mathrm{H_2O_2}$ oxidation increases by $3557-7560\%$, while $\mathrm{O_3}$ driven sulfate decreases by $55-75\%$. 
The overall SU enhancement is thus dominated by the $\mathrm{H_2O_2}$ pathway.
Increased cloud droplet numbers from additional sulfate aerosol result in significantly enhanced radiative cooling effect and stronger shortwave cloud radiative forcing, decreasing global mean SWCRF from $-0.35~\rm{W\,m^{-2}}$ to $-0.65$ and $-0.97~\rm{W\,m^{-2}}$ for bulk and surface-specific OA acidity, respectively.
These estimates are significant compared to the current uncertainty range for global effective radiative forcing from 
aerosol--cloud interactions ($-1.7$ to $-0.3~\rm{W\,m^{-2}}$) \cite{forster:ipcc:2021}.


Interestingly, surface-specific suppressed acid dissociation can lead to even stronger responses of SU, CDNC, and radiative cooling, than bulk OA acidity. 
Surface-specific OA acid dissociation increases the global SU burden, whereas regional burdens (Table~\ref{tbl:SU}) and distribution across CCN relevant accumulation and coarse modes (Supplementary information Table~\ref{tab:combined_modes_deltas}) may either increase or decrease, and both CDNC (Table~\ref{tbl:cdnc}) and SWCRE (Table~\ref{tbl:SWCRE}) are overall enhanced, relative to the bulk acidity condition.
These results of the fully coupled climate model differ from our previous box model simulations \cite{sengupta:acp:2024, sengupta:asnt:2024}, where suppression of OA acid dissociation consistently reduced SU, CDNC, and SWCRE, compared to the bulk acidity condition. 
Box models resolve aerosol microphysics, but lack coupling to large-scale atmospheric processes, where transport, mixing, and meteorological feedbacks dynamically shape aerosol populations and influence cloud formation potential.
The climate effects of both bulk and surface-specific OA acid dissociation are spatially heterogeneous, reflecting strong interactions between aerosol composition, water availability, and cloud dynamics in the fully coupled model. 
The atmosphere is a highly complicated, or even complex, system and manifestations of molecular-level processes on the global scale may result in emergent phenomena that cannot be predicted solely from responses in isolated conditions.


Surface-specific effects on cloud droplet formation and aerosol radiative effects have previously been investigated in terms of aqueous OA surface activity.
Surface adsorption of OA can decrease aqueous surface tension and promote cloud droplet activation, but the simultaneous depletion of the interior bulk in small droplets with large surface-area-to-volume ratio may strongly reduce the overall effect \cite{prisle:acp:2021, prisle:accounts:2024}. 
\citet{prisle:grl:2012} showed that global CDNC increased by $24\%$ and SWCRE 
by $-1.27~\mathrm{W \, m^{-2}}$ (compared to a total SWCRE$=-46.7~\mathrm{W \, m^{-2}}$) when droplet surface tension was reduced according bulk OA properties.
Contrary to the present work, the corresponding changes were only $1\%$ (CDNC) and $-0.08$~$\mathrm{W \, m^{-2}}$ (SWCRE) when surface-specific effects were also accounted for. 
This highlights the large potential effects of specific OA aqueous-phase properties on aerosol chemistry and microphysics, but also substantially different estimates of atmospheric impacts between considerations of surface-specific and well-known bulk aqueous conditions.



The overall acid dissociation behavior of surface-active OA in aqueous droplets likely results from both bulk and surface-specific conditions, with relative contributions depending on droplet size and OA composition, concentration, and surface activity.
For submicron droplets with large surface-area-to-volume ratios, surface effects may dominate, while OA in larger droplets may behave more like in bulk solutions \cite{bzdek:pnas:2020}.
Fine-mode aerosols ($\rm <1~\mu m$) are typically more acidic than coarse-mode ($\rm >2.5~\mu m$) \cite{pye:acp:2020, Kakavas:acp:2021}. 
In smaller size fractions, where existing low pH can suppress further acid dissociation of OA, the influence on aerosol chemistry and cloud properties may therefore resemble the surface-specific conditions considered here.
This is particularly relevant in regions with frequent new particle formation, such as over land, where aerosols tend to remain in fine-modes and experience the highly acidic conditions typical of urban and industrial areas \cite{ruan:gmd:2022, Kakavas:acp:2021}. 
In contrast, marine and remote continental regions often exhibit lower sulfate concentrations and higher aerosol pH \cite{Kakavas:acp:2021}, allowing OA acid dissociation to play a more significant role in promoting pH-dependent chemistry and aerosol--cloud--interactions.


Our results highlight that continued development of process-level constraints, particularly those that distinguish surface-specific from bulk-phase effects and incorporate pH-dependent OA parameterizations, are key to improving model fidelity in aerosol--cloud--climate interactions.
Furthermore, they show the importance of monitoring global aerosol acidity, even in regions where it is not a primary environmental concern, as aerosol acidity may still play a significant role in climate dynamics.

\section{Methods}\label{sec11}

We implement OA acid dissociation in the global climate model ECHAM-HAMMOZ (version ECHAM6.3.0-HAM2.3-MOZ1.0) following the method of \citet{sengupta:acp:2024} and \citet{sengupta:asnt:2024}.
The implementation considers the effect of concentration-dependent OA acid dissociation on both aerosol chemistry and water activity as described in the following sections.

\subsection {OA acid dissociation in aerosol chemistry}
When OA acid dissociation is not considered ($\rm{K_{a}}=0$), the default $\rm{H^+}$ concentration in ECHAM is denoted by $\rm{[H^+]_{0}}$ and obtained from water and aqueous phase sulfate concentrations as
\begin{equation}\label{Hconc}
\left[\rm{H^+}\right]_{\rm{0}}=\left[\rm{H^+}\right]_{\rm{initial}}+\frac{\left[\rm{SU}\right]}{\rm{LWC}\times \rm{MW}_{\rm{SO_4^{2-}}}},
\end{equation} 
where $\left[\rm{H^+}\right]_{\rm{initial}}$ is the hydrogen ion concentration calculated from the cloud pH, $\rm{LWC}$ $\rm{[kg\,m^{-3}]}$ is the cloud liquid water content, $\rm{MW}_{\rm{SO_4^{2-}}}$ is the molar weight of the sulfate anion, 
and the soluble sulfate concentration $\left[\rm{SU}\right]$ is obtained from the summation of soluble sulfate mass in all sizes.

We introduce OA acid dissociation by modifying eq. \ref{Hconc} to obtain the total hydrogen ion concentration in the aerosol population as
\begin{equation}\label{Hconc_modified}
\left[\rm{H^+}\right]_{\rm tot}=\left[\rm{H^+}\right]_{\rm{initial}}+\frac{\left[\rm{SU}\right]}{\rm{LWC}\times \rm{MW}_{\rm{SO_4^{2-}}}}+\left[\rm{H^+}\right]_{\rm{HA}},
\end{equation}
where $\rm{[H^+]_{HA}}$ is the concentration of the hydrogen ions dissociated by the OA acid (denoted as HA). Considering partial deprotonation of the OA acid, the acid dissociation degree $\alpha$ is given by
\begin{equation}
\label{diss2}
    \alpha=\frac{[\rm{A^-}]}{\rm{[HA]_{tot}}} = \frac{\rm{[H^+]}_{\rm{HA}}}{\rm{[HA]_{tot}}},
\end{equation}
where $\rm{[HA]_{tot}}$ is the total concentration of the organic acid derived from OA mass fraction.

For a highly diluted solution (e.g., $\rm{[HA]_{tot}} < 0.001$ $\rm{mol\,L^{-1}}$, which is significantly lower than measured organic acid concentrations in marine environments \cite{mochida:jgratmospheres:2002}), we assume the mole fraction-based mean activity coefficient $\gamma_{\pm}^{2}=1$. Under these conditions, the relationship between $\alpha$ and the acid dissociation constant $\rm{K_a}$ is
\begin{equation}\label{alpha}
    \alpha=\frac{-\rm{K_a}+\sqrt{\rm{K^2_a}+4\rm{K_a}\times \rm{[HA]_{tot}}}}{2\rm{[HA]_{tot}}}.
\end{equation}
We assume that all OA is acidic and has properties corresponding to decanoic acid, a commonly observed surface-active atmospheric fatty acid \cite{khwaja:atmenv:1995, mochida:jgratmospheres:2002, prisle:acp:2010, prisle:acp:2012, vepsalainen:acp:2023}. 
We therefore use the bulk $\mathrm{pK_a}$ of decanoic acid from the literature to describe the bulk-phase acid dissociation behavior of OA in aqueous aerosols. 
We denote this bulk value as $\mathrm{pK_a^B}$, and for decanoic acid, $\mathrm{pK_a^B} = 4.9$ \cite{martell:springer:1974}.
Using $\mathrm{pK_a^B}$ and $\rm{[HA]_{tot}} $, we calculate the concentration of dissociated hydrogen ions from OA, $[\mathrm{H}^+]_{\mathrm{HA}}$, and include it in the total aerosol-phase hydrogen ion concentration, $[\mathrm{H}^+]_{\mathrm{tot}}$, as defined in equations~\ref{diss2} and~\ref{Hconc_modified}. This acidity further influences aqueous-phase chemistry, including sulfate formation, and cloud activation processes in the model.

\subsubsection {Sulfate formation from $\rm{H_2O_2}$ and $\rm{O_3}$ oxidation of $\rm{SO_2}$}

We then modify the sulfur chemistry module, as described in our previous work \cite{sengupta:acp:2024, sengupta:asnt:2024}, to include the effects of OA acid dissociation in the calculation of aqueous phase secondary sulfate concentration $\rm{[SO_4^{2-}]}^{\prime\prime}$. We consider that in the aerosol population, secondary sulfate is formed from the oxidation of $\rm{SO_2}$ by $\rm{H_2O_2}$ and $\rm{O_3}$ in the aqueous phase.

In an aqueous environment, $\rm{SO_2}$ exists in the bisulfite form ($\rm{HSO_3^-}$) as
\begin{equation}\label{bisul}
     \rm{SO_2} + \rm{H_2O} \rightleftharpoons \rm{HSO_3^-} + \rm{H^+}.
\end{equation}
The bisulfite anion reacts with $\rm{H_2O_2}$ through the mechanism
\begin{subequations} \label{H2O2simple}
\begin{equation}\label{eqa}
     \rm{HSO_3^-} + \rm{H_2O_2} \rightleftharpoons \rm{HOOSO_2^-} + \rm{H_2O},
\end{equation}
\begin{equation}\label{eqb}
     \rm{HOOSO_2^-} + \rm{H^+} \rightleftharpoons \rm{HOOSO_2H},
\end{equation}
\begin{equation}\label{eqc}
     \rm{HOOSO_2H} \rightarrow 2\rm{H^+} + \rm{SO_4^{2-}}.
\end{equation}
\end{subequations}
The reaction rate for this $\rm{H_2O_2}$ oxidation pathway can be written as 
\begin{equation}\label{ap1}
\frac{\partial}{\partial t}\left[\rm{SO_4^{2-}} \right]^{\prime\prime}=\frac{k_4\left[\rm{H_2O_2}\right]\left[\rm{SO_2}\right]}{\left[\rm{H^+}\right]_{\rm{tot}}+0.1}
\end{equation} 
where $\rm{[H^+]_{tot}}$ is calculated using eq. \ref{Hconc_modified}, and the rate constant $k_4$ is calculated by
\begin{equation}\label{ap2}
k_4=8\times{10}^4\exp{\left(-3650\left(\frac{1}{T}-\frac{1}{298}\right)\right)},
\end{equation} 
where $T$ is the cloud temperature. Equation \ref{ap1} is $\rm{pH}$ insensitive \cite{liu:pnas:2020}. 
Equations~\ref{eqb} and~\ref{eqc} represent the current standard mechanism used in climate models for aqueous-phase sulfate production by $\rm{H_2O_2}$ oxidation, and therefore, we use this mechanism as the base case scenario in our study. In this base case, OA acid dissociation is not included ($\rm{K_a} = 0$), and the aqueous secondary sulfate concentration is calculated using the $\rm{pH}$-insensitive rate expression given in Eq.~\ref{ap1}.

For the scenarios that include OA acid dissociation, we calculate $\rm{[SO_4^{2-}]}^{\prime\prime}$ from the $\rm{H_2O_2}$ oxidation pathway using the formulation by \citet{liu:pnas:2020}, which is valid for $\rm{pH} > 2$. In this framework, the OA acid (denoted by HA) is treated as a weak acid that can act as a general proton donor. We then apply the general acid catalysis mechanism, replacing the standard reaction steps given in Eqs.~\ref{eqb} and~\ref{eqc} with

\begin{equation}
\label{ps1}
\rm{HOOSO_2^-} + \rm{HA} \rightarrow 2\rm{H^+} + \rm{SO_4^{2-}} + \rm{A^-}.
\end{equation}
The rate expression for $\rm{[SO_4^{2-}]}^{\prime\prime}$ formation from this mechanism is
\begin{equation}\label{nsulh2o2}
\frac{\partial}{\partial t}\left[\rm{SO_4^{2-}}\right]^{\prime\prime}=\left(k+\frac{k_{\rm{HA}}[\rm{HA}]}{[\rm{H^+}]_{\rm{tot}}}\right)\rm{K_{a1}}[\rm{H_2O_2}][\rm{SO_2}],
\end{equation} 
where $\rm{K_{a1}}$ is the first acid dissociation constant of $\rm{H_2SO_3}$ and $k$ is a constant derived from the reaction rate coefficient and the thermodynamic equilibrium constants. $k_{\rm{HA}}$ is the overall rate constant for the general acid catalysis mechanism approximated by $\log k_{\rm{HA}}= -0.57(\rm{pK_a})+6.83$ \cite{liu:pnas:2020}. 
This approximation for $k_{\rm{HA}}$ as a function of the $\mathrm{pK_a}$ of the OA acid (HA) was derived by \citet{liu:pnas:2020} for an ionic strength of $I = 0.5\,\mathrm{mol\,kg^{-1}}$. We therefore assume the same ionic strength for aqueous droplets in all our calculations.

$\rm{O_3}$ reacts with 
$\rm{HSO_3^-}$ in the aqueous phase according to
\begin{equation}
\label{eqo3}
\rm{HSO_3^-} + \rm{O_3} \rightarrow \rm{H^+} + \rm{SO_4^{2-}} + \rm{O_2}.
\end{equation}
The secondary sulfate concentration from the $\rm{O_3}$ oxidation is then given by 
\begin{equation}\label{ap5}
\frac{\partial}{\partial t}\left[\rm{SO_4^{2-}}\right]^{\prime\prime}=\left(k_{51}+\frac{k_{52}}{\left[\rm{H^+}\right]_{\rm{tot}}}\right)\left[\rm{O_3}\right]\left[\rm{SO_2}\right],
\end{equation}
where rate constants $k_{51}$ and $k_{52}$ are calculated from
\begin{equation}\label{ap6}
k_{51}=4.39\times{10}^{11}\exp{\left(\frac{-4131}{T}\right)}
\end{equation} 
and 
\begin{equation}\label{ap7}
k_{52}=2.56\times{10}^3\exp{\left(\frac{-996}{T}\right)}.
\end{equation}

\subsubsection {Secondary sulfate mass in the aerosol module}

The secondary sulfate concentration from $\rm{H_2O_2}$ and $\rm{O_3}$ oxidation of $\rm{SO_2}$ is calculated using Eqs.~\ref{nsulh2o2} and \ref{ap5}, respectively. These concentrations are used to compute the secondary sulfate mass, $m\left({\rm{SO_4^{2-}}}\right)^{\prime\prime}_t$, in the aqueous sulfur chemistry module. The sulfur chemistry module is coupled to the aerosol microphysical module such that, at each time step $t$, the total sulfate mass is given by

\begin{equation} \label{massSU}
m{\left({\rm{SU}}\right)}_t = m{\left({\rm{SU}}\right)}_0 + m\left({\rm{SO_4^{2-}}}\right)^{\prime\prime}_t,
\end{equation}
\\
where $m{\left({\rm{SU}}\right)}_0$ is the initial mass of SU from the emission inventory and $m\left({\rm{SO_4^{2-}}}\right)^{\prime\prime}_t$ is the secondary sulfate mass calculated in the sulfur chemistry module at time $t$. For each time step, $m{\left({\rm{SU}}\right)}_t$ is calculated in the aerosol module using the $m\left({\rm{SO_4^{2-}}}\right)^{\prime\prime}_t$ obtained from the aqueous sulfur chemistry module. Then $m{\left({\rm{SU}}\right)}_t$ is used in the calculation of the aerosol microphysical processes, which includes nucleation, condensation, coagulation and hydration processes, in each time step.
\subsection{OA acid dissociation in water activity}
\label{vhfsec}
Water activity in aqueous aerosols is a measure of the availability of water in the system relative to pure water and is influenced by the amount of solute in the aerosol. It is defined as the ratio of the partial vapor pressure of water in the solution to the vapor pressure of pure bulk water under the same ambient conditions. 

In the ECHAM climate model, the default amount of solute, without considering OA acid dissociation, is given by
\begin{equation} 
\label {ns_default} 
n_s = i_{\rm{SU}} n_{\rm{SU}} + i_{\rm{OA}} n_{\rm{OA}} + i_{\rm{SS}} n_{\rm{SS}}, \end{equation}
where $n_{\rm{SU}}$, $n_{\rm{OA}}$, and $n_{\rm{SS}}$ represent the initial moles of soluble sulfate, organic aerosol, and sea salt, respectively, which are obtained from the emission inventory. $i_{\rm{SU}}=3$, $i_{\rm{OA}}=1$  and $i_{\rm{SS}}=2$ are the corresponding van't Hoff factors, such that sulfate and sea salt are considered to be fully dissociated, while OA is considered as undissociated (fully protonated, $\rm{K_{a}}=0$).

We consider the effects of concentration-dependent partial OA acid dissociation in the calculation of $n_s$ by modification of two parameters. Firstly, the amount of soluble sulfate at time $t$, $n_{\rm{SU},t}$, is calculated using $m{\left({\rm{SU}}\right)}_t$ obtained from eq. \ref{massSU}. Secondly, the van't Hoff factor for organic acid dissociation, $i_{\rm{OA}}$, is modified to reflect partial acid dissociation of the OA. 
We calculate $i_{\rm{OA}}$ from the dissociation degree, $\alpha$ (from eq. \ref{alpha}), using 
\begin{equation} \label {vhfactor}
		i_{\rm{OA}} = 1+\alpha(n_{\rm{ions}}-1),
\end{equation}
where $n_{\rm{ions}}$ is the fractional number of ions formed from one 
molecule of the organic acid.
The total available amount of solute at time $t$, $n_{s,t}$, is then calculated for simulations considering OA acid dissociation as
\begin{equation} 
\label {ns}
n_{s,t}=i_{\rm{SU}}n_{\rm{SU},t}+\left(1+\alpha(n_{\rm{ions}}-1)\right)n_{\rm{OA}}+i_{\rm{SS}}n_{\rm{SS}}.
\end{equation}
$n_s$ is modified at each time step by introducing OA acid dissociation according to $i_{\rm{OA}}$ from eq. \ref{vhfactor} and modified sulfate mass from eq. \ref{massSU}. 

\subsection{Surface-specific OA acid dissociation}

In addition to the bulk-phase OA acid dissociation, we also implement an empirical representation of the surface-specific shift in acid--base protonation equilibrium, as observed in recent X-ray photoelectron spectroscopy (XPS) experiments \cite{prisle:acp:2012, ohrwall:jphyschem:2015, werner:pccp:2018}. These highly surface-sensitive experiments showed that, across a wide range of solution pH, the degree of dissociation at the aqueous surface differs systematically from that in the bulk. For a range of atmospheric organic acids and bases, including decanoic acid and other simple mono-carboxylic acids, the equilibrium at the surface is strongly shifted toward the neutral species. This suppression of organic acid dissociation corresponds to an apparent increase in $\mathrm{pK_a}$ of approximately 1--2 pH units, compared to bulk solution values.

We assume, as a first approximation, that the entire OA fraction is characterized by the surface-specific dissociation behavior. Specifically, we assume that the suppressed acid dissociation observed for surface-adsorbed organic acids applies uniformly throughout the droplet. This assumption is supported by thermodynamic calculations showing that surface-active organic compounds, including Suwannee River Fulvic Acid (SRFA), Nordic Aquatic Fulvic Acid (NAFA), sodium octanoate, sodium decanoate, sodium dodecanoate, and sodium dodecyl sulfate, strongly partition to the surface during much of the droplet growth and activation process \cite{prisle:acp:2010, prisle:acp:2021, vepsalainen:acp:2023, prisle:accounts:2024}. This effect is especially important in microscopic and submicron-sized droplets, where the surface-area-to-volume ratio is large. For example, spherical droplets with diameters of $D_{\rm{wet}} = 0.1$, $1$, and $10$~$\mu$m have surface-area-to-volume ratios of $60$, $6$, and $0.6$~$\mu\mathrm{m}^{-1}$, respectively \cite{prisle:acp:2010, prisle:accounts:2024}.

To represent the surface-specific suppressed acid dissociation, we shift the $\mathrm{pK_a}$ from the well-known bulk-phase value $\mathrm{pK_a^B}$ by 1--2 pH units, such that the apparent $\mathrm{pK_a}$ are $\mathrm{pK_a^{S1}} = \mathrm{pK_a^B} + 1$ and $\mathrm{pK_a^{S2}} = \mathrm{pK_a^B} + 2$, representing moderate and strong suppression, respectively, of acid dissociation in the surface. The intrinsic bulk $\mathrm{pK_a}$ of each organic compound is not changed in our implementation. Rather, the dissociation response to a given pH is modified to reflect the observed surface-specific dissociation behavior. We calculate the dissociation degree $\alpha$ from Eq.~\ref{alpha} using each apparent $\mathrm{pK_a}$, and use Eq.~\ref{vhfactor} to calculate the corresponding van’t Hoff factors $i_{\rm{OA}}$.

\subsection{ECHAM-HAMMOZ atmosphere model}

We used the ECHAM-HAMMOZ global climate model (version ECHAM6.3.0-HAM2.3-MOZ1.0). The main atmospheric component is ECHAM 6 
which includes the sub-model HAM2.3 for tropospheric aerosols 
and a sub-model MOZ for trace-gas chemistry. 
We used the T63 spectral truncation for the horizontal grid, with 47 vertical levels which follow the terrain and use the hybrid vertical coordinate representations 
The configurations for the relevant parameters used in this work are given in Table \ref{tbl:1}.

\begin{table}
  \caption{Parametrizations used with ECHAM6.3-HAM2.3 for the simulations}
  \label{tbl:1}
  \begin{tabular}{ll}
    \hline
    Parameter  &  Model configuration  \\
    \hline
    Activation Scheme   & Lohmann etal., 1999 \cite{lohmann:jgratmos:1999}, Abdul-Razzak \& Ghan 2000 \cite{abdul-razzak:jgratmos:2000}   \\
    Shortwave and longwave radiation & Rapid Radiative Transfer Model (RRTM)  \\
    Aerosol module  & Modal M7  \\
    Vertical layers & 47 \\
    Domain resolution  & 1.87$^{\circ}$ latitude x 1.87$^{\circ}$ longitude  \\
    Simulation time & 1999-2003  with 3-month spin-up time \\
    Anthropogenic emission inventory & CMIP5 \\
    \hline
  \end{tabular}
\end{table}

Cloud droplet number concentration (CDNC) burden, sulfate (SU) burden, and cloud liquid water content (LWC) were obtained directly from the ECHAM output variables {\it atmospheric burden of CDNC}, {\it atmospheric burden of sulfate}, and {\it vertically integrated cloud water}, respectively. The resulting shortwave radiative effect and forcing predicted for each OA acid dissociation condition were obtained as described in the following.

\subsubsection{Shortwave cloud radiative effect and forcing}
We calculated the shortwave cloud radiative effect (SWCRE) as the difference between all-sky and clear-sky shortwave radiative fluxes at the top of the atmosphere for each OA acid dissociation condition, following \citet{tang:acp:2020}:

\begin{equation}
\mathrm{SWCRE} = \mathrm{srad0} - \mathrm{sraf0}.
\end{equation}
\\
Here, {\it srad0} and {\it sraf0} are ECHAM output fields representing the net shortwave flux under all-sky and clear-sky conditions, respectively.

To quantify the impact of OA acid dissociation on radiative forcing, we then computed the shortwave cloud radiative forcing (SWCRF) as the change in SWCRE relative to a clean, pre-industrial baseline as

\begin{equation}
\mathrm{SWCRF} = \mathrm{SWCRE}{\text{present-day}} - \mathrm{SWCRE}{\text{pre-industrial}}.
\end{equation}
\\
The pre-industrial reference simulation used anthropogenic emissions from the Community Emissions Data System (CEDS) for the year 1750 \cite{hoesly:gmd:2018}.

\subsection{Data analysis}

We analyzed all diagnostics as 5-year medians using monthly ECHAM output from $1999-2003$. We calculated normalized median differences ($\Delta \mathrm{X}$) for each model output variable $\mathrm{X}$ with respect to a chosen reference condition $\mathrm{Y}$ as
\begin{equation}
\Delta \mathrm{X} \big|_{\mathrm{Y}} = \frac{\mathrm{X}_{\mathrm{pK_a}}(\overline{i,j}) - \mathrm{X}_{\mathrm{Y}}(\overline{i,j})}{\mathrm{X}_{\mathrm{Y}}(\overline{i,j})},
\label{eq:deltaX}
\end{equation}
where $\mathrm{Y}$ denotes the reference scenario, either $\mathrm{K_{a}}=0$ (no OA acid dissociation) or $\mathrm{pK^{B}_a}$ (bulk OA acid dissociation), depending on the comparison. The indices $\mathrm{i}$ and $\mathrm{j}$ represent latitude and longitude in the model grid, and the overline indicates a 5-year median value.

\subsubsection{Significance test}
\label{Sec:sig}
We assessed the significance of differences between model simulations using the Wilcoxon signed-rank test, a non-parametric method for analyzing matched-pair data \cite{woolson:book:2005}. The Wilcoxon signed-rank test is defined by 
\begin{equation}
W = \sum_{i=1}^{N} \rm{sgn}(x_i - y_i) \cdot R_i,
\label{eqn 5}
\end{equation}
where \(W\) is the resulting test statistic,
 \(N\) is the total number of paired observations, and
\(x_i\) and \(y_i\) represent the \(i\)-th element in vectors \(x\) and \(y\), respectively.
The function $\mathrm{sgn}$ returns $-1$ if the argument is negative, $0$ if the argument is zero, and $1$ if the argument is positive.
\(R_i\) represents the rank of the absolute difference $\left| x_i - y_i \right|$ when sorted in ascending order.

We compute the test using paired monthly medians of a given variable (e.g., CDNC or SU) between two scenarios, such as $\rm{pK^{B}_a}$ vs. the base-case $\rm{K_{a}}=0$ conditions. The null hypothesis assumes a median difference of zero. The test statistic \(W\) is used to compute a p-value, which reflects the probability of observing the data under the null hypothesis.
To reflect varying degrees of statistical confidence, we use three thresholds: $\rm{p}<0.01$ (strong evidence), $\rm{p}<0.05$ (moderate evidence), and $\rm{p}<0.1$ (weak evidence). These thresholds are visualized in plots using white circles of different sizes, with larger circles corresponding to lower p-values, and thus stronger evidence of a given result being different from the null hypothesis.

\backmatter









\section{Supplementary Information}
\subsection{Sulfate distribution across aerosol modes}

We explored the distribution of sulfate across aerosol size modes, i.e. Aitken (dry particle radius $R_p=0.005{-}0.05\,\mu$m), accumulation ($R_p=0.05{-}0.5\,\mu$m), and coarse ($R_p>0.5\,\mu$m), under different OA acid dissociation conditions. 
Table~\ref{tab:combined_modes_deltas} presents the predicted global median sulfate mass fraction, $X_\mathrm{Sulfate}$ (\%), defined as the fraction of sulfate mass relative to the total mass of all species (sulfate, OA, sea salt, black carbon, dust) in each size mode. Sulfate mass in each aerosol size mode is obtained directly from the ECHAM model output variable \textit{M\_SO4\_<mode>}. 
The atmospheric total column sulfate mass burden SU shown in Fig.~1 in the main text is obtained by summing \textit{M\_SO4\_<mode>} over all aerosol size modes and integrating vertically. In Table~\ref{tab:combined_modes_deltas}, we also 
list the total aerosol number concentration, $N_\mathrm{Total}$ (m$^{-3}$), and the mass-fraction weighted equivalent sulfate number concentration, $N_\mathrm{Sulfate}= N_\mathrm{Total} \times X_\mathrm{Sulfate}$ (m$^{-3}$), for each size mode.
$N_\mathrm{Sulfate}$ provides an estimate of how the sulfate mass burden contributes to the total particle number concentration in each mode.

\begin{table}[h]
\centering
\caption{
Global median sulfate mass fractions, $X_\mathrm{Sulfate}$ (\%), total aerosol number concentrations, $N_\mathrm{Total}$ (m$^{-3}$), and mass-fraction weighted equivalent sulfate number concentrations, $N_\mathrm{Sulfate}$ (m$^{-3}$), in different aerosol size modes predicted under different OA acid dissociation conditions.
}
\label{tab:combined_modes_deltas}
\begin{tabular}{lcccc}
\toprule
\textbf{Parameter} & $K_\mathrm{a} = 0$ & $\mathrm{pK_a^B}$ & $\mathrm{pK_a^{S1}}$ & $\mathrm{pK_a^{S2}}$ \\
\midrule
\multicolumn{5}{c}{\textbf{Aitken mode} ($R_p = 0.005$--$0.05\,\mu$m)} \\
$X_\mathrm{Sulfate}$ (\%)               & 51.2  & 52.1  & 52.3  & 56.0 \\
$N_\mathrm{Total}$ ($\mathrm{m^{-3}}$)  & $1.80\!\times\!10^8$ & $2.57\!\times\!10^8$ & $2.25\!\times\!10^8$ & $1.98\!\times\!10^8$ \\
$N_\mathrm{Sulfate}$ ($\mathrm{m^{-3}}$)& $0.92\!\times\!10^8$ & $1.34\!\times\!10^8$ & $1.18\!\times\!10^8$ & $1.11\!\times\!10^8$ \\
\midrule
\multicolumn{5}{c}{\textbf{Accumulation mode} ($R_p = 0.05$--$0.5\,\mu$m)} \\
$X_\mathrm{Sulfate}$ (\%)               & 5.3   & 5.6   & 5.5   & 5.2 \\
$N_\mathrm{Total}$ ($\mathrm{m^{-3}}$)  & $4.58\!\times\!10^7$ & $6.15\!\times\!10^7$ & $6.03\!\times\!10^7$ & $5.97\!\times\!10^7$ \\
$N_\mathrm{Sulfate}$ ($\mathrm{m^{-3}}$)& $0.25\!\times\!10^7$ & $0.35\!\times\!10^7$ & $0.33\!\times\!10^7$ & $0.31\!\times\!10^7$ \\
\midrule
\multicolumn{5}{c}{\textbf{Coarse mode} ($R_p > 0.5\,\mu$m)} \\
$X_\mathrm{Sulfate}$ (\%)               & 4.8   & 5.7   & 6.0   & 7.4 \\
$N_\mathrm{Total}$ ($\mathrm{m^{-3}}$)  & $2.35\!\times\!10^5$ & $3.11\!\times\!10^5$ & $3.46\!\times\!10^5$ & $3.87\!\times\!10^5$ \\
$N_\mathrm{Sulfate}$ ($\mathrm{m^{-3}}$)& $0.11\!\times\!10^5$ & $0.18\!\times\!10^5$ & $0.21\!\times\!10^5$ & $0.29\!\times\!10^5$ \\
\bottomrule
\end{tabular}
\end{table}

For all OA acid dissociation conditions, the Aitken mode consistently shows the highest and the coarse mode the lowest $N_\mathrm{Total}$, $X_\mathrm{Sulfate}$, and $N_\mathrm{Sulfate}$, between the different size modes. For the accumulation mode, covering the size range most relevant for cloud droplet growth and activation adding to CDNC, $N_\mathrm{Total}$, $X_\mathrm{Sulfate}$, and $N_\mathrm{Sulfate}$ are roughly 2 orders of magnitude higher than for the coarse mode and 1 lower than for the Aitken mode. 
For each aerosol size mode, including OA acid dissociation ($\mathrm{pK_a^B}$) increases predicted sulfate mass fractions, $X_\mathrm{Sulfate}$, total aerosol number concentrations, $N_\mathrm{Total}$, and mass-fraction weighted equivalent sulfate number concentrations, $N_\mathrm{Sulfate}$, compared to the no OA acid dissociation condition ($K_\mathrm{a} = 0$). Additional sulfate mass from considering OA acid dissociation therefore increases aerosol number concentrations across all size modes.
In contrast, surface-specific dissociation conditions ($\mathrm{pK_a^{S1}}$ and $\mathrm{pK_a^{S2}}$) further increase $X_\mathrm{Sulfate}$ for the Aitken and coarse modes, but decrease $X_\mathrm{Sulfate}$ for the accumulation mode, compared to the bulk condition ($\mathrm{pK_a^B}$).
Concurrently, $N_\mathrm{Sulfate}$ and $N_\mathrm{Total}$ decrease for both Aitken and accumulation modes, but increase further for the coarse mode, for surface-specific compared to bulk OA acid dissociation conditions.
Surface-specific conditions therefore lead to significantly increased sulfate mass and number concentrations of highly CCN active coarse mode aerosol, but simultaneously decrease both sulfate and total number concentrations in the much more abundant Aitken and accumulation modes. The CCN activity of aerosol populations in the latter size range in particular are highly sensitive to ambient conditions, leading to strong potential effects of interactions with the local environment.

\subsection{Cloud liquid water content}

\begin{figure}[H]
  \centering
     \includegraphics[width=\textwidth]{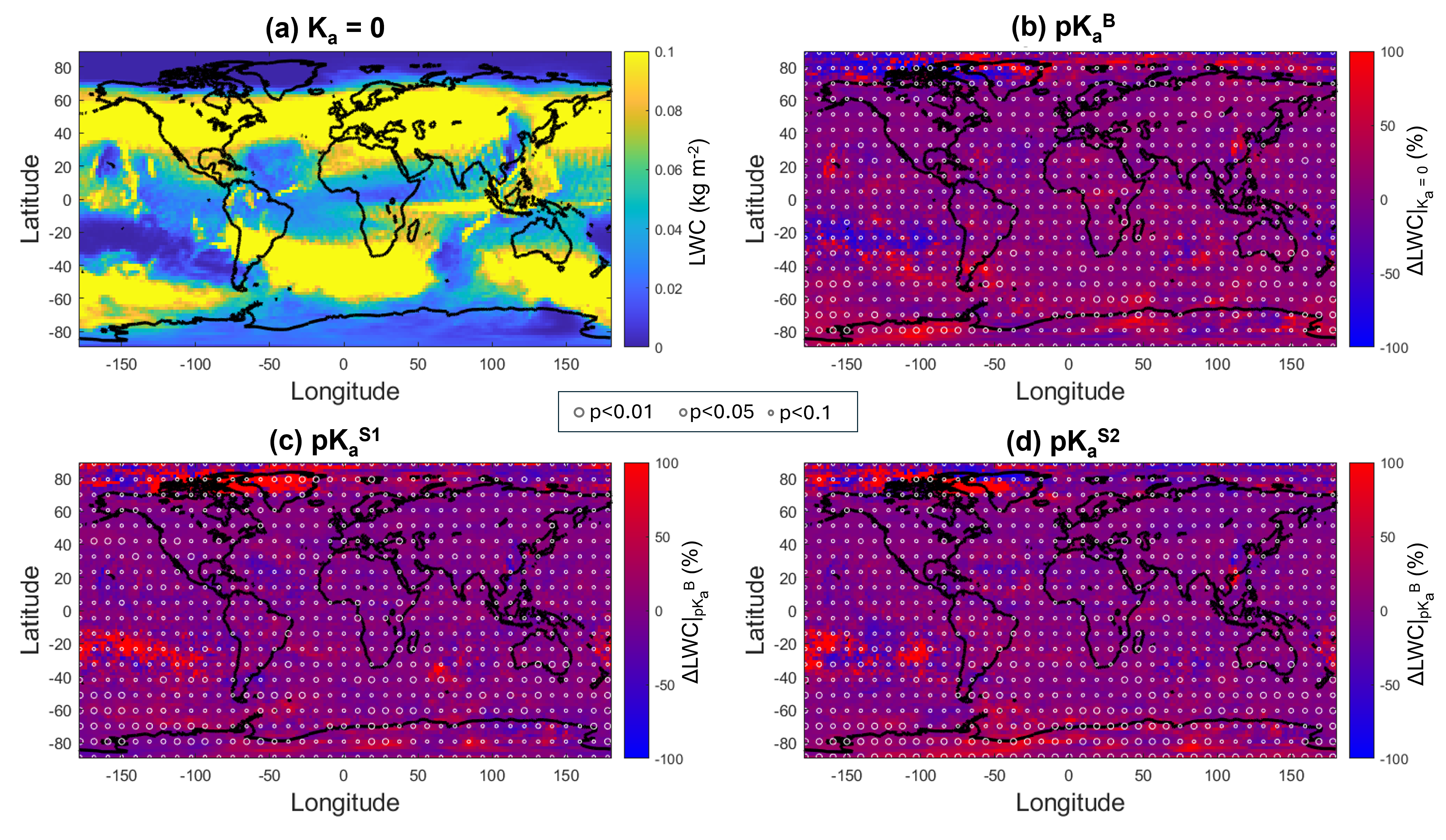}
      \caption{Cloud liquid water content (LWC) burden shown as (a) the total column burden for the no OA acid dissociation condition ($\rm K_a=0$) as 5-year ($1999-2003$) medians in absolute units (\(\rm kg \, m^{-2}\)), and the column burden differences (b) with respect to $\rm K_a=0$ ($\Delta \mathrm{LWC} \big|_{\mathrm{K_{a}=0}}$, \%) for bulk OA acid dissociation ($\mathrm{pK^{B}_a}$), and with respect to $\mathrm{pK^{B}_a}$ ($\Delta \mathrm{LWC} \big|_{\mathrm{pK_a^B}}$, \%) for the surface-specific OA acid dissociation (c) $\mathrm{pK^{S1}_a}$ and (d) $\mathrm{pK^{S2}_a}$. Statistical significances of differences given as $p < 0.01$ (strong), $p < 0.05$ (medium), or $p < 0.1$ (weak) are represented by large, medium, and small circles, respectively.}
  \label{lwc}
\end{figure}

Figure \ref{lwc} shows the global distribution of total column cloud liquid water content (LWC) as 5-year medians from \(1999\)--\(2003\) and its sensitivity to different conditions of OA acid dissociation. Panel (a) shows the absolute LWC column burden (\(\rm kg \, m^{-2}\)) for the condition of no OA acid dissociation (\(\rm K_a=0\)). 
Panel (b) shows the change in LWC column burden \(\Delta \mathrm{LWC} \big|_{\mathrm{K_{a}=0}}\) for the bulk acid dissociation condition $\mathrm{pK^{B}_a}$ relative to \(\rm K_a=0\), while panels (c) and (d) show \(\Delta \mathrm{LWC} \big|_{\mathrm{pK_a^B}}\) for surface-specific suppressed acid dissociation ($\mathrm{pK^{S1}_a}$ and $\mathrm{pK^{S2}_a}$, respectively) relative to $\mathrm{pK^{B}_a}$. 
The statistical significance of the differences in panels (b), (c), and (d) is evaluated as strong ($p < 0.01$, large circles), moderate ($p < 0.05$, medium circles), and weak ($p < 0.1$, small circles) using the Wilcoxon signed-rank test.  
Average LWC column burdens for each OA acid dissociation condition, both globally and regionally over land and ocean areas, are summarized in Table \ref{tbl:lwc}.

Globally, including OA bulk acid dissociation ($\mathrm{pK_a^B}$, panel a) increases LWC from $0.067$ to $0.071$~$\rm{kg\,m^{-2}}$ ($5.86\%$) compared to \(\rm K_a=0\). 
Regionally, the enhancement (\(\Delta \mathrm{LWC} \big|_{\mathrm{K_{a}=0}}\)) is stronger over land ($7.27\%$) than over oceans ($3.64\%$). 
These increases are consistent with the concurrent rise in CDNC (Fig.~1 in the main text).
The surface-specific OA acid dissociation conditions ($\mathrm{pK_a^{S1}}$ and $\mathrm{pK_a^{S2}}$) lead to further increases in LWC compared to the bulk condition. 
For $\mathrm{pK_a^{S1}}$, \(\Delta \mathrm{LWC} \big|_{\mathrm{pK_a^B}}= 0.56\%\) (\(\Delta \mathrm{LWC} \big|_{\mathrm{K_{a}=0}}=6.52\%\)), with land regions showing a \(\Delta \mathrm{LWC} \big|_{\mathrm{pK_a^B}}= 0.54\%\). 
For the strongest surface suppression of acid dissociation ($\mathrm{pK_a^{S2}}$), \(\Delta \mathrm{LWC} \big|_{\mathrm{pK_a^B}}= 1.83\%\), with a $2.60\%$ increase over land and a modest $0.85\%$ increase over oceans, compared to the bulk OA acidity condition. 

\begin{table}[h]
\centering
\caption{5-year absolute median LWC burden ($\rm kg\,m^{-2}$) and relative changes (\%) in LWC burden with respect to $\mathrm{K_{a}}=0$ ($\Delta \mathrm{LWC} \big|_{\mathrm{K_{a}=0}}$) and $\mathrm{pK_a^B}$ ($\Delta \mathrm{LWC} \big|_{\mathrm{pK_a^B}}$) for bulk ($\mathrm{pK_a^B}$) and surface-specific ($\mathrm{pK_a^{S1}}$, $\mathrm{pK_a^{S2}}$) OA acid dissociation over the entire globe (world), land, and ocean regions.} 
\label{tbl:lwc}
\begin{tabular}{llccc}
\toprule
Region & $\mathrm{pK_a}$ & LWC ($\rm kg\,m^{-2}$) & $\Delta \mathrm{LWC} \big|_{\mathrm{K_{a}=0}}$ ($\%$) & $\Delta \mathrm{LWC} \big|_{\mathrm{pK_a^B}}$ ($\%$) \\ 
\midrule
\multirow{4}{*}{World} 
    & $\mathrm{K_{a}}=0$    & $0.0670$ & --    & --    \\
    & $\mathrm{pK_a^B}$     & $0.0710$ & $5.86$  & --    \\
    & $\mathrm{pK_a^{S1}}$  & $0.0714$ & $6.52$  & $0.56$ \\
    & $\mathrm{pK_a^{S2}}$  & $0.0723$ & $7.86$  & $1.83$ \\
\midrule
\multirow{4}{*}{Land}  
    & $\mathrm{K_{a}}=0$    & $0.0611$ & --    & --    \\
    & $\mathrm{pK_a^B}$     & $0.0655$ & $7.27$  & --    \\
    & $\mathrm{pK_a^{S1}}$  & $0.0658$ & $7.66$  & $0.54$ \\
    & $\mathrm{pK_a^{S2}}$  & $0.0672$ & $9.95$  & $2.60$ \\
\midrule
\multirow{4}{*}{Ocean} 
    & $\mathrm{K_{a}}=0$    & $0.0792$ & --    & --    \\
    & $\mathrm{pK_a^B}$     & $0.0821$ & $3.64$  & --    \\
    & $\mathrm{pK_a^{S1}}$  & $0.0829$ & $4.75$  & $1.00$ \\
    & $\mathrm{pK_a^{S2}}$  & $0.0828$ & $4.59$  & $0.85$ \\
\bottomrule
\end{tabular}
\end{table}

\subsection{Cloud droplet number concentration}

\begin{figure}[H]
  \centering
     \includegraphics[width=\textwidth]{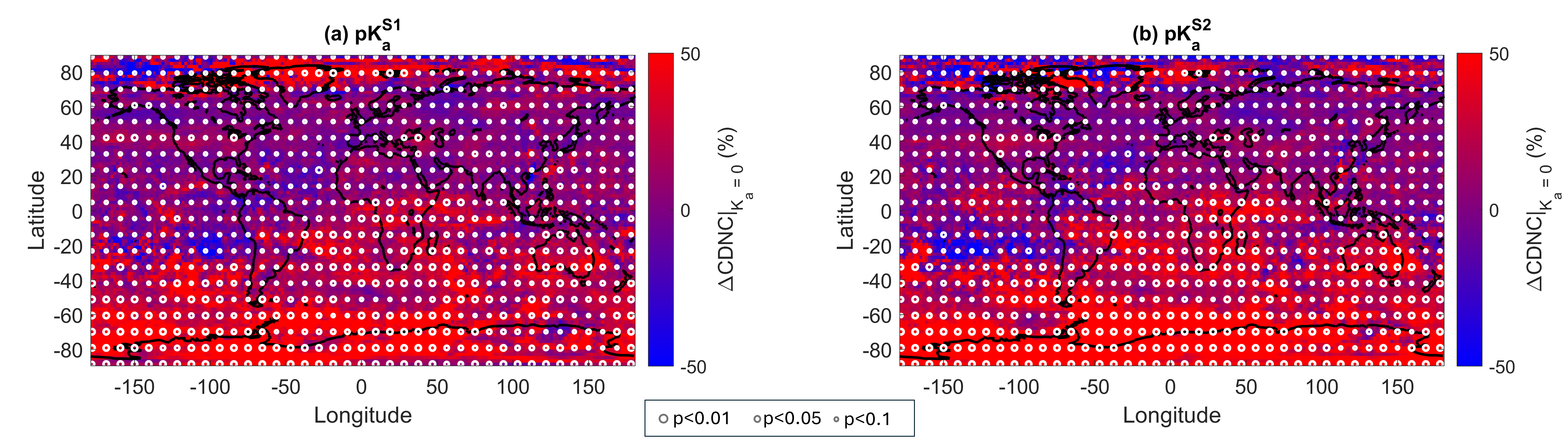}
      \caption{CDNC column burden differences with respect to $\rm K_a=0$ ($\Delta \mathrm{CDNC} \big|_{\mathrm{K_{a}=0}}$, \%) for the surface-specific OA acid dissociation  $\mathrm{pK^{S1}_a}$ (panel a) and $\mathrm{pK^{S2}_a}$ (panel b). Statistical significance (white circles) is indicated as large ($p < 0.01$), medium ($p < 0.05$), and small ($p < 0.1$).}
  \label{cdncKa0}
\end{figure}

Figure \ref{cdncKa0} shows the change in CDNC column burden ($\Delta \mathrm{CDNC} \big|_{\mathrm{K_{a}=0}}$) for surface-specific suppressed acid dissociation conditions, $\mathrm{pK^{S1}_a}$ (panel (a) and $\mathrm{pK^{S2}_a}$ (panel b), with respect to the $K_\mathrm{a} = 0$ condition.
Both scenarios show widespread positive $\Delta \mathrm{CDNC} \big|{\mathrm{K_{a}=0}}$, with larger increases for $pK^\mathrm{S2}_a$ than the $pK^\mathrm{S1}_a$ condition. Statistically significant changes ($p < 0.01$) occur predominantly over land and in mid-latitude marine regions.
These patterns are consistent with the $\Delta \mathrm{CDNC} \big|_{\mathrm{pK_a^B}}$ shown in the main text (Fig.~1, panels c and d), where both surface-specific conditions exhibit additional enhancement in CDNC compared to the bulk acid dissociation condition.



\bmhead{Acknowledgements}

The authors warmly thank Dr. Harri Kokkola, Finnish Meteorological Institute, and Irfan Muhammed, University of Eastern Finland, for valuable support on running ECHAM-HAMMOZ. 
This project has received funding from the European Research Council (ERC) under the European Union's Horizon 2020 research and innovation program, project SURFACE (grant agreement no. 717022). The authors also gratefully acknowledge the financial contribution from the Research Council of Finland, including grant nos. 308238, 314175, and 335649.



\section*{Declarations}


\begin{itemize}
\item Funding: This project has received funding from the European Research Council (ERC) under the European Union's Horizon 2020 research and innovation program, project SURFACE (grant agreement no. 717022). The authors also gratefully acknowledge the financial contribution from the Research Council of Finland, including grant nos. 308238, 314175, and 335649.
\item Conflict of interest/Competing interests: The authors declare no competing interests.
\item Ethics approval and consent to participate: Not applicable.
\item Consent for publication: Not applicable.
\item Data availability: The model output data generated in this study are available at [Zenodo], [DOI/link to be added upon acceptance].  
\item Materials availability: Not applicable.
\item Code availability: The ECHAM-HAMMOZ model is available upon request to the HAMMOZ consortium (https://redmine.hammoz.ethz.ch).  
\item Author contribution: GS implemented the modifications in the model with contributions from KG and PRJ. GS performed the simulations and calculations, with contributions from KG and PRJ. GS, KG, and NLP analyzed the results. GS and NLP wrote the manuscript. NLP conceived, planned, and supervised the project and secured funding. All authors reviewed and approved the final manuscript.
\end{itemize}

\bibliography{sn-bibliography}

\end{document}